# Valuing life detection missions

Edwin S. Kite* (University of Chicago), Eric Gaidos (University of Hawaii),
Tullis C. Onstott (Princeton University).   * kite@uchicago.edu

Recent discoveries imply that Early Mars was habitable for life-as-we-know-it (Grotzinger et al. 2014); that Enceladus might be habitable (Waite et al. 2017); and that many stars have Earth-sized exoplanets whose insolation favors surface liquid water (Dressing & Charbonneau 2013, Gaidos 2013). These exciting discoveries make it more likely that spacecraft now under construction – Mars 2020, ExoMars rover, JWST, Europa Clipper – will find habitable, or formerly habitable, environments. Did these environments see life? Given finite resources (\$10bn/decade for the US[1]), how could we best test the hypothesis of a second origin of life? Here, we first state the case for and against flying life detection missions soon. Next, we assume that life detection missions will happen soon, and propose a framework (Fig. 1) for comparing the value of different life detection missions:

$$\text{Scientific value} = (\text{Reach} \times \text{grasp} \times \text{certainty} \times \text{payoff}) / \$ \qquad (1)$$

After discussing each term in this framework, we conclude that scientific value is maximized if life detection missions are flown as hypothesis tests. With hypothesis testing, even a nondetection is scientifically valuable.

**Should the US fly more life detection missions?**

Once a habitable environment has been found and characterized, life detection missions are a logical next step. Are we ready to do this?

*The case for emphasizing habitable environments, not life detection:* Our one attempt to detect life, Viking, is viewed in hindsight as premature or at best uncertain. In-space life detection experiments are expensive. Other expensive experimental disciplines, such as US laser fusion and US particle physics, have histories that are cautionary tales about over-promising. Today, the search for life beyond Earth sustains Congressional and public enthusiasm for planetary science. This enthusiasm could die down if life detection missions yield nondetections (even if they are false negatives). Perhaps the real payoff would be something so unexpected that it would be missed. To the extent that the science questions cannot be precisely defined in advance (Heng 2016), a better motivation for planetary missions is pure exploration – to push the boundaries of what humans can do, visit, and know. This argues that the next generation of astrobiology missions should emphasize detecting and characterizing habitable environments, rather than the search for extinct or extant life.

*The case for flying more life detection missions:* Life appears near the start of Earth's geologic record and could be widespread in the Universe. A detection of a second origin of life has the potential to transform the science of biology. It would also provide guidance about our own future (Bostrom 2008), including the human role in the Solar System. If we indefinitely defer decisive life detection tests, then the search for life is simply PR for planetary science and astronomy. To optimally spend the $10 bn allotted to us over the next decade, astrobiologists should aim to test for life as quickly, as decisively, and as often as possible.

---

[1] We pick \$10bn/decade as a rough estimate of current US spending on astrobiology. We note that international cooperation gathers talent, brings a reduced probability of cancellation, and is valuable in itself.

As the number of habitable extraterrestrial environments increases, the arguments for developing life detection missions that target those locations become stronger. Recent developments, such as the publication of the report of the Science Definition Team for the Europa Lander (Hand 2017), show that life detection missions are again being seriously considered. Therefore, we now need a framework for valuing different life detection mission concepts.

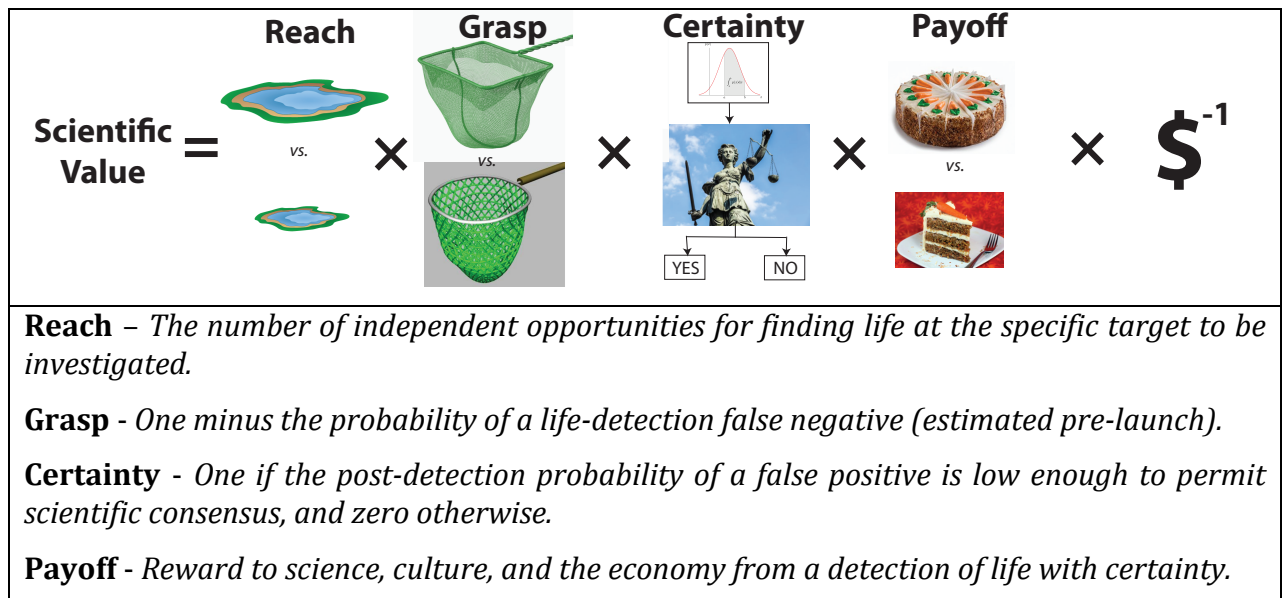

**Reach** – *The number of independent opportunities for finding life at the specific target to be investigated.*

**Grasp** - *One minus the probability of a life-detection false negative (estimated pre-launch).*

**Certainty** - *One if the post-detection probability of a false positive is low enough to permit scientific consensus, and zero otherwise.*

**Payoff** - *Reward to science, culture, and the economy from a detection of life with certainty.*

**Figure 1.** Framework for valuing life detection missions.

Here we emphasize science goals, not the specifics of mission implementation. A rigorous comparison of any two specific missions would have to consider many nuances to the design of the specific missions – for example instrument capabilities, trajectory design, and risk versus cost – and we do not attempt to do that here.

**Reach:**

One measure of value is a mission's *reach* - how many independent opportunities for finding life are there at the specific target to be investigated? This depends on (1) the size and diversity of the target environment, and (2) the fraction of the target environment that is effectively sampled by the mission.

The odds that life-as-we-know-it will emerge and persist get better the greater the area of rock-water interfaces, the greater the dynamical cross-section for panspermia, and the longer-lived the habitable conditions. More diverse environments are more likely to encompass the (unknown) conditions required for life to establish itself. By this logic, looking for microbes on a globally habitable Early Mars may offer better odds than looking for life in now-frozen impact-generated lakes of liquid water on Titan (Artemieva & Lunine 2003). On the other hand, Earth-size planets can remain geologically active for so long as to erase ancient fossils (Earth and Venus are examples; Sleep & Bird 2007). The potential reach of exoplanet missions is ≳$10^8$ habitable-zone worlds, although we do not yet know what fraction of habitable-zone planets are habitable, and so do not yet know how the number of habitable targets depends on telescope specifications (e.g. Postman et al. 2010).

Reach is maximized when geological processes allow a single mission to sample for life that evolved in a voluminous environment. For this to happen, life or biosignatures must be conveyed to the probed location by groundwater flow, cryovolcanism, winds, or ocean currents. Winds and currents swiftly mix planet-sized environments. Therefore, the reach of a mission that probes an atmosphere, the surface of a globally habitable planet, or an ocean is large. Exoplanet spectra also probe global environments. By contrast, groundwater flow can be slow and spatially restricted (petroleum can be trapped, and rarely flows >100 km from its source, for a similar reason; Hunt 1995). The transport of living cells can be still more restricted than the transport of soluble biosignatures. Yet subsurface environments may be longer-lived than surface environments (e.g. Mars; Grimm et al. 2017) and can preserve life's signatures well (e.g. petroleum again; Peters et al. 2005). Nonetheless, a mission searching for a rock-hosted biosphere has a reach that is a small fraction of the planet's crustal habitable volume. Proving or falsifying the hypothesis that a rock-hosted biosphere exists deep beneath a hostile-to-life surface in any given planetary crust may simply be too expensive for the current budget.

The scale of an environment is a crude yardstick for its probability of hosting life. For example, Europa has ~100× the seafloor area of Enceladus. Intuitively, Europa might have a higher probability for life (all other things equal), in part because larger environments are more likely to be stable and persist[2] – but not 100× more. This intuition can be captured by using a log prior (or a log log prior; Lacki 2016). These priors say that, if our ignorance about the likelihood of the origin of life spans very many orders of magnitude – which it does – then it is likely that either suitable planet-sized environments are almost all inhabited (habitability is all that's required, and life is inevitable), or almost all uninhabited (life as chemical accident). It is rather unlikely that (say) ~50% are inhabited, because there is no reason for the scale of the environment needed for origin of life to be equal to the scale of a planet, even to order of magnitude (Carter 1983, Lacki 2016). Priors that behave in this way moderate the importance of reach. With a log prior, the more ignorant we are, the more it makes sense to look for life in habitable environments that are small, short-lived, or both. Examples of such environments include asteroid parent bodies (Gaidos & Selsis 2006), and rock bodies on Earth that were buried to uninhabitable depths but have since been exhumed (Onstott 2016). For the same reason, a search for life on the hundred closest habitable planets is not much less valuable than a sweep of the entire Galaxy.

**Grasp and Certainty:**

We define grasp as equal to one minus the probability of a life-detection false negative (estimated pre-launch). Both planetary processes and instrumental effects can degrade biosignatures, and so both contribute to the false negative probability. False-negative probability is defined relative to the best-available pre-launch understanding of the distribution of biosignatures in the specific materials to be actually investigated (based on Earth's geologic record, Earth-analog environments, lab work on biomarker preservation, e.t.c). A mission that asks the wrong question (relative to our pre-launch model of what is there to be found) has low grasp regardless of whether it is ready to successfully answer that question.

---

[2] But see Fuller et al. (2016) for a mechanism by which Enceladus' ocean could be as long-lived as Europa's ocean.

We define certainty as equal to unity if the post-detection probability of a false positive is low enough to permit scientific consensus, and zero otherwise. Recent examples of scientific certainty include the discoveries of Neanderthal DNA (Krings et al. 1997), gravitational waves (Abbott et al. 2011), and the $^{16}$O enrichment of the Sun relative to the planets (McKeegan et al. 2011). Solid scientific progress demands certainty. Certainty is maximized by integrating multiple approaches (Schulze-Makuch et al. 2015). For example, by combining molecular, isotopic, and textural clues, scientific certainty about ancient microbial life is possible (Wacey et al. 2009, Knoll et al. 2016). By contrast, consensus is elusive for isotopes-only claims and texture-only claims for Hadean life (Bell et al. 2015, Nutman et al. 2016). The histories of claims about Precambrian life and life in Martian meteorites (McKay et al. 1996) illustrate that reaching consensus involves a loop: analyze, interpret, critique, then analyze again. This loop takes time. Time may be in short supply on an in-situ mission (Hand 2008).

Returned-sample life detection, therefore, has better certainty and grasp than in-situ life detection. Sample return allows thorough molecular (e.g. Summons et al. 2008), isotopic (e.g. Stephan et al. 2016) and textural inspection, but only up to the limits set by sample size and by contextual documentation at the sampled site. If the returned samples are too small or too few in number (for example due to an overestimate of the in-space density of interesting samples; Westphal et al. 2014), then a false negative can occur. Nevertheless, sample return to Earth maximizes grasp and certainty (Mustard et al. 2013). This comes at a cost: for sample return from a habitable body, >$1bn. This cost is increased by rules that are set by NASA. Given constraints on spacecraft mass and cost, the coring and sample caching payload on a mission that is intended to be the first in a sample return campaign (e.g. Mars 2020) might seem to maximize future certainty (from subsequent sample return) but at the expense of instruments that could increase grasp – unless the in-situ analyses are able to complement the package by increasing grasp (Mustard et al. 2013). Certainty for extant-life detection comparable to that of sample return might be achieved by combining multiple proxies for life, such as motility or perhaps consumption of redox gradients, that can be measured in-situ (Nadeau et al. 2016, Weiss et al. 2000). Some of these in-situ proxies may be unmeasurable in a returned sample. The grasp of in-situ instruments will improve with further technology investments. However, relative to Earth laboratories, flight instruments have lower certainty for life detection and cannot be swapped out nor upgraded in response to initial results.

Despite great reach, exoplanet life detection using inner-Solar-System telescopes has low certainty. Spectroscopic detection of high levels of $O_2$ and/or chemical disequilibrium have been proposed as exoplanet biomarker candidates (Schwietermann et al. 2017, Krissansen-Totton et al. 2018). Both are really bio-hinters, because most detectable gas combinations can be produced without life. Exceptions, such as isoprene or CFCs, are too rare in Earth's atmosphere for detection at interstellar distances (Seager et al. 2012). Unfortunately, the $CH_4+O_2$ combination yielded by Earth's biosphere has been undetectable in long-range transmission spectroscopy throughout Earth's history: today because of low $CH_4$ and refraction (Misra et al. 2014), and in the Precambrian due to low $CH_4$ (Reinhard et al. 2017) or negligible $O_2$. If a large fraction of planets have both abundant $CH_4$ and abundant $O_2$, then this would be hard to explain in terms of abiotic transients (Catling & Kasting 2017, Krissansen-Totton et al. 2018). However, we have no reason to think that a large fraction of inhabited planets will be so cooperative. Non-gas biosignatures such as the vegetation red edge (Seager et al. 2005) are intriguing, but for these, little effort has yet been spent on modeling to identify false positives.

These problems cannot be side-stepped by probabilistic approaches, because our prior uncertainty on life's abundance is so broad (Lacki 2016), and rocky planets are diverse. If we want to do a Bayesian model comparison of with-life versus no-life models (given some exoplanet data), then we need to know the probability of the data given the no-life model (Catling et al. 2017). This requires a forward model for atmospheric evolution on uninhabited yet habitable planets. Although it is easy to build such a model on a computer, our modest predictive power for Solar System atmosphere composition suggests humility about predictions for exoplanet atmospheres that are potentially much more diverse (Zahnle & Catling 2017). Moreover, the models are in danger of being over-fit to a few Solar System data points. It is risky to use uninhabitable rocky exoplanets as the no-life control set, because abiotic false positives are correlated with some abiotic processes that promote habitability. Moreover, Earth may be too limited a template for an inhabited planet due to anthropocentric selection effects. Thus, although we might test the hypothesis that biospheres are "infrequent" (they might stand out with respect to other habitable but uninhabited planets), we cannot deal with the case that biospheres are "very uncommon" (sample size will always be insufficient to both detect the very-uncommon biosphere itself and also to rule out equally uncommon, but expected, abiotic false-positive scenarios), nor the "prolific" biosphere case (all or almost all habitable planets have life). Thus, we might detect a true biosignature, but not know with certainty that life is the source. For example, suppose that 99% of $O_2$-rich atmospheres have $O_2$ as the result of life. Solar System telescopes could never approach 99% certainty that even one of those atmospheres roofs a biosphere, because abiotic $O_2$ production scenarios (Schwieterman et al. 2017) cannot be ruled out to this confidence level[3]. This has implications for the use of JWST. Should we look for biosignatures around a few planets, or instead probe for habitability in a larger sample of planets (Bean et al. 2017)? If reaching certainty about exoplanet life detection requires observations of many *uninhabited* "control cases," then more planets are better.

**Payoff:**

To find an independent origin of life would be a scientific breakthrough. The breakthrough would have a payoff that would depend on the nature of the evidence. Ancient-fossil evidence would be studied using the same techniques used to study Precambrian fossils on the Earth. Depending upon the preservation, these techniques constrain metabolism, composition and cell size and structure, but say little about genetics (Knoll et al. 2016). Therefore, ancient-fossil life would have limited direct impact on sciences outside astrobiology unless the fossils preserved their molecular structure. Space-telescope detection of an exoplanet biosphere (Dalcanton et al. 2015) would offer tantalizingly little information about the organisms themselves. This might stimulate interstellar flight if the biosphere orbits a nearby star (Lubin 2016), or the construction of very large single-target space telescopes to study/monitor the biosphere and surface. Detection of extant life (or young fossils that retain DNA or equivalent) would offer the biggest payoff. For example, information about intact life might transform the biological sciences – which, via the health sector, underpin >10% of Gross World Product. Microbial life that shares a common ancestor with life-as-we-know-it might be easiest to exploit economically, but analysis of life that evolved completely independently could solve a wider range of scientific puzzles.

---

[3] At least as long as high-resolution data to constrain abiotic-planet models are confined to the Solar System.

Even a low-payoff detection would supercharge space exploration, and thus potentially speed up the discovery of high-payoff life elsewhere. This moderates the importance of payoff. Similarly, confirmed in-situ detection of a living organism (with no characterization) is almost as good as retrieval, because a retrieval mission would then be launched by one or more countries with minimal delay.

**Valuing life detection missions as hypothesis tests**

Inevitably, our notional attempt to apply the criteria of reach, grasp, certainty and payoff (Table 1) is mottled by our blind spots and prejudices as authors. Our intent is to encourage a broader discussion that draws on the community's collective expertise. Moreover, Table 1 could be reset by a scientific wildcard, such as liquid water at <1 km depth on Europa, or by a technology development, such as fission reactors for deep-space missions (McNutt et al. 2015). Nevertheless, two low-cost opportunities appear to have potential out of proportion to current funding. First, and perhaps the most compelling, is SETI. The other is study of natural origin of life experiments in Earth's subsurface – isolated water pockets that were first sterilized, then exhumed to habitable depths (Holland et al. 2013). These terrestrial environments are dwarfed by the crustal volume of Mars, but using a log prior this should not count against them too strongly. Intraterrestrial origin-of-life experiments can be investigated by sterile drilling, which is in any case a needed technology for ocean-worlds exploration. This argues for NSF-NASA or DoE-NASA cooperation.

Using origin-of-life research to drive target selection is risky. Because the geologic setting(s) of abiogenesis is (are) unknown (e.g. McCollom & Seewald 2013), geologically diverse targets – and targets with the highest production rates of free energy able to drive chemosynthesis – are the best bets. (Titan's surface might be an example of a suitably diverse target, but only if life can establish itself in non-aqueous fluids – COEL 2007, Shapiro and Schulze-Makuch 2009). However, prioritizing a mission because of any one origin-of-life hypothesis is questionable. For example, the environment targeted for life detection can be distant (physically and chemically) from the environment of abiogenesis: fragmentation during impacts early in Solar System history enables re-inoculation after giant impacts (Wells et al. 2003). Nevertheless, prebiotic systems where life did not arise might inform origin-of-life-research. Life might yet be created in the laboratory – perhaps tomorrow. While scientifically significant, would this inform the search for life on other worlds? Probably not: there may be many mechanisms for abiogenesis – many roads to life – and because of the timescale and chemical limitations of laboratory work, we should not expect the one that first works in the lab to be the same as the one that happened at planetary scale.

Once a habitable environment has been identified, refined constraints on fluxes of free energy and nutrients offer (limited) guidance for target selection. Energy and nutrient fluxes could scale with biomarker production/concentration, which when elevated offers better sensitivity for life detection. However, life endures in nutrient-poor environments (Priscu et al. 1999), many energy conservation strategies are possible (Schulze-Makuch & Irwin 2002), and – if given an initial minimal nutrient budget and an energy source – a biosphere may self-sustain via heterotrophy, recycling and adaptation.

Current reconnaissance missions, such as MRO and Europa Clipper, have a strong science return regardless of astrobiology outcome. However, life detection requires instruments that differ from those used to study habitable environments. Therefore, future Solar System astrobiology planners will have to weigh continued characterization of habitable environments against life detection.

A life detection mission is a hypothesis test if the probability of life is greatly reduced by a nondetection (Platt 1964). Missions that are not hypothesis tests – usually due to low grasp – have low value within the framework we propose here. Although it has been said that "exploration often cannot be hypothesis testing" (Chyba & Phillips 2001, Hand 2017), hypothesis-testing has served us well in the past (Mars Science Program Synthesis Group 2004). Hypothesis-testing also offers a clear basis for reallocating resources in response to negative results (Smolin 2006). Hypothesis testing is necessary but insufficient for high science value: with post-1996 data in hand, we now see that the 1976 Viking landers had both low grasp and low reach.

Recent successful missions have uncovered apparently habitable environments. As the number of known habitable environments increases, it will be tempting to rebalance the US astrobiology portfolio away from continued exploration of habitable environments, and towards testing the hypothesis of life. Each target offers unique tradeoffs. Proposed life detection missions may be valued by sizing up their reach, grasp, certainty, and payoff (e.g., Table 1). Missions that emphasize life detection should test astrobiology hypotheses. Framing good hypotheses requires precursor missions. Life detection missions have low scientific value unless a negative result can guide future decisions and future missions – for example, whether or not to move on to more promising targets.

**Acknowledgements.** We are grateful for reviews from Chris McKay, Alfonso Davila, and an anonymous reviewer. We thank Bethany Ehlmann and Chris House for feedback, without implying agreement with the points made here.

| | Mission profile | Reach | Grasp | Certainty | Payoff | New technologies needed | Cost |
|---|---|---|---|---|---|---|---|
| Space-based | Return sample of ancient Mars surface environments (e.g., Mars 2020 to Jezero) | ✓✓ | ✓ or ✓✓ | ✓✓ | ✓ | | $$$ |
| | Return sample of ancient Mars subsurface environments | – or ✓ | ✓ | ✓✓ | ✓ | | $$$ |
| | Mars deep drill, in-situ measurements only (assuming present-day aquifers exist) | – or ✓ | ✓ | ✓ | ✓✓ | Compact high-output power source | $$$$ |
| | Seek refugia on present Mars surface, in-situ measurements only | n.a. | ✓ | ✓ | ✓✓ | | $/$$ |
| | Mars in situ paleontology (e.g., ExoMars lander) | ✓ or ✓✓ | – | – or ✓ | ✓ | Enhanced by improved in-situ instruments | $$ |
| | Ocean world drill to ≲1m (e.g. Europa lander) | ✓✓ (*) | – or ✓ (*) | ✓ | ✓✓ | | $$ ($$$?) |
| | Ocean world, probe liquid water ocean in-situ | ✓✓ | ✓✓ | ✓✓ | ✓✓ | Compact high-output power source | $$$$ |
| | Ocean-sourced plume in-situ (e.g., Enceladus Life Finder) | ✓✓ | ✓ | ✓ | ✓✓ | Enhanced by improved in-situ instruments | $$ |
| | Ocean-sourced plume sample return | ✓✓ | ✓ | ✓✓ | ✓✓ | | $$ |
| | Retrieve earliest Earth materials (from "Earth's attic," the Moon) | ✓ | – | ✓✓ | – or ✓ | | $/$$ |
| | Exoplanet survey transit or direct imaging | ✓✓✓ | ✓ | – | ✓ | | $$$/$$$$ |
| | Investigate material from interstellar interloper (e.g. 'Oumuamua) | – | – | ✓ | ✓ or ✓✓ | High $\Delta V$ to land, or return samples | $$$ |
| | Interstellar probe | ✓✓ | ✓ | ✓ | ✓ | Interstellar propulsion & communication | $10^{11}$-$10^{12}$ |
| Earth based | Laboratory origin of life experiments | – | ✓✓ | ✓✓ | ✓ | | <$ |
| | Probe natural origin of life experiments on Earth | – | ✓ | ✓✓ | ✓✓ | | $ |
| | SETI | ✓✓✓ | – | ✓✓ | ✓✓✓ | | $ |

**Table 1.** A matrix for assessing the life detection case for selected potential mission profiles. The values we have assigned are notional, and our goal is to encourage a broader discussion that draws on the community's collective expertise. – = does not strengthen the life detection case for a mission. ✓ = ambivalent implications for the life detection case for a mission. ✓✓ = Bolsters the life detection case for a mission. ✓✓✓ = Offers strong support for the life detection case for a mission. The four criteria are multiplicative not additive (Fig. 1). * = Depends on geologic history of landing site. Notional costs: $ = <1 bn. $$ = 1-3 bn. $$$ = 3-10 bn. $$$$ = >10 bn

**References.**

Abbott, B.P. et al. (LIGO Scientific Collaboration and Virgo Collaboration) 2016. Observation of Gravitational Waves from a Binary Black Hole Merger, Phys. Rev. Lett. 116, 061102, arXiv:1602.03837.

Artemieva, N. & J. Lunine, 2003, Cratering on Titan: impact melt, ejecta, and the fate of surface organics, Icarus 164, 471-480.

Bean, J.L., D.S. Abbot and E. M.-R. Kempton, 2017, A Statistical Comparative Planetology Approach to the Hunt for Habitable Exoplanets and Life Beyond the Solar System, Astrophys. J. Lett. 841:L24.

Bell E.A., Boehnke P., Harrison T.M., Mao W.L., 2015, Potentially biogenic carbon preserved in a 4.1 billion-year-old zircon. Proc. Natl. Acad. Sci. 112, 14518-14521. doi:10.1073/pnas.1517557112.

Bostrom, N., 2008, Where Are They? MIT Technology Review, May/June 2008.

Carter, B., 1983, The anthropic principle and its implications for biological evolution. Philosophical Transactions of the Royal Society of London A, 310:34-363.

Catling, D.C. & Kasting, J.F., 2017, Atmospheric evolution on inhabited and lifeless worlds, Cambridge University Press, 592 pp.

Catling, D.C.; Krissansen-Totton, J.; Kiang, N.Y.; Crisp, D.; Robinson, T.D.; DasSarma, S.; Rushby, A.; Del Genio, A.; Bains, W.; Domagal-Goldman, S., 2017, Exoplanet Biosignatures: A Framework for Their Assessment, arXiv:1705.06381.

Chyba, C.F., and C.B. Phillips, 2001, Possible ecosystems and the search for life on Europa, Proc. Natl. Acad. Sci. 98, 801-804.

COEL (Committee on the Origins and Evolution of Life), 2007, The Limits of Organic Life in Planetary Systems, National Research Council, ISBN: 0-309-66906-5, 116 pp.

Dalcanton, J., et al. 2015, From Cosmic Birth to Living Earths: The Future of UVOIR Space Astronomy, arXiv:1507.04779.

Dressing, C., and Charbonneau, D., 2013, The Occurrence Rate of Small Planets around Small Stars, Astrophys. J. 767(1), article id. 95, 20 pp.

Fuller, J.; Luan, J.; Quataert, E., 2016, Resonance locking as the source of rapid tidal migration in the Jupiter and Saturn moon systems, Monthly Notices of the Royal Astronomical Society 458, 3867-3879.

Gaidos, E., 2013, Candidate Planets in the Habitable Zones of Kepler Stars, Astrophys. J. 770, article id.90

Gaidos, E., & F. Selsis, 2006, From Protoplanets to Protolife: The Emergence and Maintenance of Life, p.929-944 *in* B. Reipurth, D. Jewitt, and K. Keil (Eds.), Protostars & Planets V, University of Arizona Press.

Grimm, R.E.; Harrison, K.P.; Stillman, D.E.; Kirchoff, M.R., 2017, On the secular retention of ground water and ice on Mars, Journal of Geophysical Research: Planets 122, 94-109.

Grotzinger, J. et al. 2014, A Habitable Fluvio-Lacustrine Environment at Yellowknife Bay, Gale Crater, Mars, Science 343, 6169, id. 1242777.

Hand, E., 2008, Mars exploration: Phoenix: a race against time, Nature 456, 690-695.

Hand, K.P., Murray, A. E.; Garvin, J., et al., 2017, Report of the Europa Lander Science Definition Team.

Heng, K., 2016, The Imprecise Search for Extraterrestrial Habitability, American Scientist, 104(3), 146-, doi:10.1511/2016.120.146.

Holland, G., Lollar, B.S., Li, L., Lacrampe-Couloume, G., Slater, G. F., and Ballentine, C. J., 2013, Deep fracture fluids isolated in the crust since the Precambrian era, Nature 497, 357-360.

Hunt, J., 1995, Petroleum geochemistry and geology, 2$^{nd}$ edn., W.H. Freeman, 743 pp.

Kasting, J.F.; Kopparapu, R.; Ramirez, R.M.; Harman, C.E., 2014, Remote life-detection criteria, habitable zone boundaries, and the frequency of Earth-like planets around M and late K stars, Proc. Natl. Acad. Sci. 111, 12641-12646

Knoll A.H., Bergmann K.D., Strauss J.V., 2016, Life: the first two billion years. Phil. Trans. R. Soc. B 371: 20150493. http://dx.doi.org/10.1098/rstb.2015.04939